\documentclass[twocolumn,preprintnumbers,amsmath,amssymb]{revtex4-1}
\usepackage{graphicx}
\usepackage{ulem}
\usepackage[svgnames]{xcolor}
\usepackage{bm}
\usepackage{multirow}
\usepackage{float}
\usepackage{titlesec}

\begin{document}

\title{Graphite in 90 T: Evidence for Strong-coupling Excitonic Pairing}

\author{Zengwei Zhu$^{1,2,*}$, Pan Nie$^{1}$, Beno\^{\i}t Fauqu\'{e}$^{3,4}$, Ross D. McDonald$^{2}$, Neil Harrison$^{2}$ and Kamran Behnia$^{1,4}$}

\affiliation{(1) Wuhan National High Magnetic Field Center and School of Physics, Huazhong University of Science and Technology,  Wuhan  430074, China.\\
(2) MS-E536, NHMFL, Los Alamos National Laboratory, Los Alamos, New Mexico, 87545, USA.\\
(3) JEIP,  USR 3573 CNRS, Coll\`ege de France, PSL Research University, 11, place Marcelin Berthelot, 75231 Paris Cedex 05, France.\\
(4)Laboratoire de Physique Et Etude des Mat\'{e}riaux (UPMC-CNRS), ESPCI Paris, PSL Research University 75005 Paris, France.\\}

\date{\today}

\begin{abstract}
Strong magnetic field induces at least two phase transitions in graphite beyond the quantum limit where many-body effects are expected. We report on a study using a state-of-the-art non-destructive magnet allowing to attain 90.5 T at 1.4 K, which reveals a new field-induced phase and evidence that the insulating state destroyed at 75 T is an excitonic condensate of electron-hole pairs. By monitoring the angle dependence of in-plane and out-of-plane magnetoresistance, we distinguish between the role of cyclotron and Zeeman energies in driving various phase transitions. We find that, with the notable exception of the transition field separating the two insulating states, the threshold magnetic field for all other transitions display an exact cosine angular dependence. Remarkably, the threshold field for the destruction of the second insulator (phase B) is temperature-independent with no detectable Landau-level crossing nearby. We conclude that the field-induced insulator starts as a weak-coupling spin-density-wave, but ends as a strong-coupling excitonic insulator of spin-polarized electron-hole pairs.
\end{abstract}

\maketitle

The quantizing effect of high magnetic fields upon the motion of electrons perpendicular to the field orientation effectively reduces the electronic dimensionality of metals. Ultimately a sufficiently strong magnetic field confines electrons to their lowest Landau level, dramatically increasing the electronic degeneracy at the Fermi energy. In a 2D system, the fractional quantum Hall effect\cite{Stormer} emerges in this regime, due to electron-electron interactions. In 3D systems, on the other hand, electrons can still move freely parallel to the applied field. The quasi-one-dimensionality of their energy spectrum leaves them vulnerable to various instabilities like charge-density-wave (CDW), spin-density-wave (SDW), excitonic insulator (EI) and valley-density-wave(VDW)\cite{Halperin,MacDonald}. As a result, the behavior of the three-dimensional(3D) electron gas in this regime has attracted tremendous attention both experimentally and theoretically \cite{Halperin,MacDonald,Celli1965,Lee1969,VDW1987}, albeit with only a limited number of experimental systems with sufficiently low carrier density to reach this limit.

In graphite, beyond its quantum limit (QL) induced by a moderate field applied along the c-axis,  several phase transitions have been observed\cite{Yaguchi,FauqueBook}. In TaP\cite{Jia2017TaP} and TaAs\cite{Ramshaw2017TaAs}, the dominant anomalies in resistivity have been attributed to the field-induced annihilation of Weyl nodes reminiscent of the total evacuation of Dirac valleys in bismuth\cite{Zhu2017bismuth}. However, only in graphite there is compelling thermodynamic evidence\cite{Benoit2017} associated with field-induced activation gaps\cite{Benoit2013}, which point to collective electronic phenomena beyond the single-particle picture.

The first experimental observation of an abrupt increase in magnetoresistance around 25 T at low temperature was reported in 1981\cite{Tanuma}. The dome-like phase diagram in the temperature-field plane was established by the study of in-plane magnetoresistance(I$\perp$c). Around 54 T, the insulating state is destroyed\cite{Yaguchi}. There are several proposals for the identity of the field-induced order, ranging from CDW\cite{Yoshioka, ArnoldCDW,Shindou1}, SDW\cite{Takada,SDWTakahashi}, EI\cite{Akiba, ZhuEI} and spin nematic excitonic insulator\cite{Pan}. Recently, a new transition and a second dome in the temperature-field plane ending at 75 T was established by out-of-plane (I$\parallel$c) measurements\cite{Benoit2013}. Signatures of these transitions in ultrasonic attenuation and velocity\cite{Benoit2017} (up to 65 T) and in Nernst coefficient \cite{Benoit2011} in fields below 45 T have been detected. Exfoliation was also used to study the thickness dependence of the transitions\cite{Thickness2018}.

Here, we present a study of  magnetoresistance in a non-destructive magnetic fields up to 90.5 T revealing an additional phase and bringing new insight to the identity of the previously-known field-induced states. The contrasting roles of orbital and Zeeman energies can be elucidated by the orientation of the applied field owing to their differing anisotropies.The $\alpha$ and $\beta$ transitions (marking the beginning of the in-plane and out-of-plane insulating response) do not deviate from a cosinusoidal angular dependence over the entire measurement range. By contrast to all other threshold fields, the 54 T transition ($\alpha'$) does not follow a cosinusoidal behavior, in contrast to all other threshold fields. This implies that the Zeeman energy plays a central role in the transition from one field-induced insulator to the other one. The combination of temperature dependence and angle dependence of the $\zeta$ transition lead us to identify phase B as a spin-polarized strong-coupling excitonic insulator (likely paring between electron and hole bands with the opposite spin) destroyed by strong magnetic field. The C-phase established afterwards displays a metallic behavior irrespective of the orientation of the charge flow (parallel and perpendicular) respective to the graphene planes. This highlights the peculiarity of the phase B where a magnetoresistanceless in-plane metallicity coexists with out-of-plane activation. Furthermore, the absence of semimetal to full semiconductor gap at the $\zeta$ transition indicates that in phase C the lowest spin-polarized hole and electron levels are still occupied.

Natural graphite samples were commercially obtained. In-plane magnetoresistance was measured with a standard four-contacts set-up. Out-of-plane magnetoresistance was measured using two pairs of electrodes attached to the top and bottom of a sample, as illustrated in the inset of the Fig.1. In both cases, the magnetic field was predominantly oriented along the c-axis of the sample, with $\theta$ is defined to be the angle between the field orientation and c-axis. High-field magnetoresistance measurements were performed both in WHMFC, Wuhan and NHMFL, los Alamos. The results were found to be consistent among several samples. In WHMFC, the current was applied by a NI-5402 signal generator worked set to 100 kHz and voltage was recorded by a NI-5105 high-speed digitizer worked at 4 MHz. A digital phase lock-in method was used to extract the magnetoresistance. High magnetic field measurements up to 90.5 T were performed at NHMFL, los Alamos. The 100 T magnet consists an inner and outer magnet. The outer magnet is driven by a generator to produce  field between 0 and 37 T at 3 s total width , followed by a faster (15 ms) capacitor bank driven pulse to 90.5 T. The Nernst effect was measured by the standard one-heater-two-thermometer method in a dilution refrigerator equipped with a piezo-rotator in a 17 T superconducting magnet on highly orientated pyrolytic graphite (HOPG) and kish graphite samples\cite{ZhuNernstGraphite,SM}.

Fig.1(a) shows field-dependence of in-plane $\rho_{xx}$ and out-of-plane $\rho_{zz}$ magnetoresistance, up to 90.5 T at 1.4 K. $\rho_{zz}$ rises by one order of magnitude at the $\beta$ transition. It drops before rising again in the vicinity of 54 T, signaling the existence of a second insulating phase named B \cite{Benoit2013}.  It drops again at 75 T and becomes relatively flat afterward upon the destruction of the phase B.  The in-plane resistivity, $\rho_{xx}$, presents a kink at the $\alpha$ transition and a small enhancement at the $\beta$ transition, becomes flat and then shows a dramatic rise above 75 T. Fig. 1(b) displays the field-dependence of anisotropy ratio $\rho_{xx}$/$\rho_{zz}$. The three field-induced cascading phases can be clearly delineated: the phase A refers to the one between the $\alpha$ and $\alpha'$ anomalies, the phase B exists between $\alpha'$ and $\zeta$ anomalies and the phase C starts at $\zeta$ and continues up to the highest explored magnetic field. The zero-field anisotropy ratio of the natural graphite sample studied here was found to be around 0.002, comparable to previous reports\cite{NautreAnisotropy,SM} Upon the application of the magnetic field, $\rho_{xx}$/$\rho_{zz}$ steadily increases and attains 0.3 at 7.5 T above which the quantum limit is reached. Then it decreases steeply following the $\beta$ and $\alpha'$ transition, recovering a more quasi-2D behavior. Finally, following the $\zeta$ transition, the system tends to become quasi-3D again. We note that different types of graphite present different anisotropy ratios\cite{Benoit2013}, and even samples of the same type can be different in their zero-field anisotropies\cite{SM}. In spite of these differences, however, the evolution of anisotropy with magnetic field was found to be reproducible \cite{SM}.

%
\begin{figure}
\includegraphics[width=9cm]{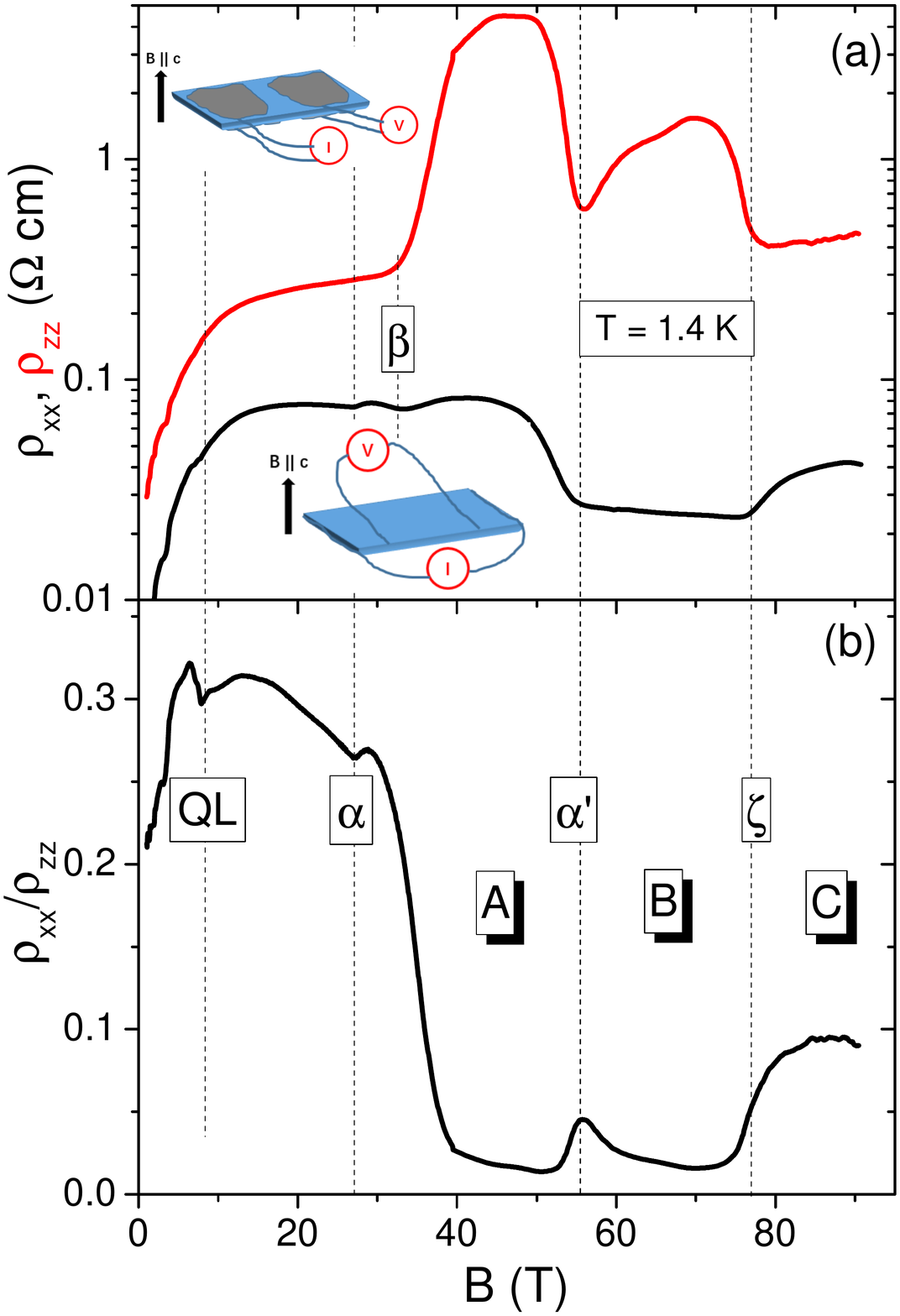}
\caption{ (a). Field-dependence of in-plane magnetoresistance $\rho_{xx}$ and out-of-plane magnetoresistance $\rho_{zz}$ up to 90.5 T at 1.4 K. The sketches illustrate configurations. (b) The field-dependence of ratio $\rho_{xx}$/$\rho_{zz}$. The quantum limit (QL) and the three phases (A, B and C) are designated.}
\end{figure}

\begin{figure}
\includegraphics[width=9cm]{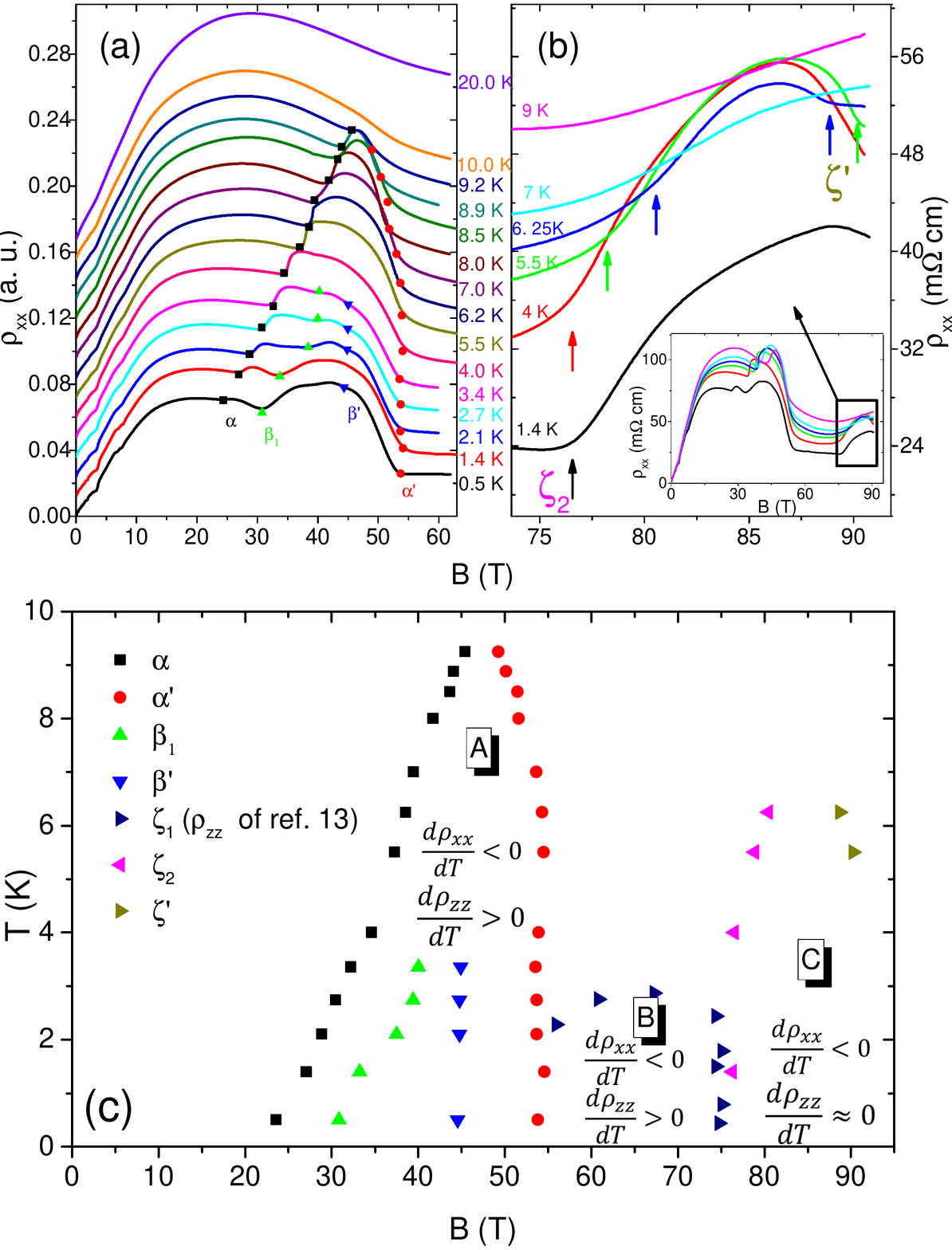}
\caption{ (a). The field-dependence of $\rho_{xx}$ at various temperatures up to 65 T. Curves are shifted for clarity. The different transitions are also labelled accordingly. (b). In-plane magnetoresistance $\rho_{xx}$ (not shifted) between 73 T and 90.5 T while the inset shows magnetoresistance in whole field range at various temperatures. The arrows shows the transitions labelled as $\zeta$ and $\zeta'$. (c).  The complete phase diagram obtained from the current work. A new phase C emerges above 75 T.  }
\end{figure}

To map out the complete phase diagram, in addition to measuring $\rho_{xx}$ up to 65 T with a regular pulsed magnet( Fig. 2 (a)), we measured $\rho_{xx}$ up to 90.5 T at different temperatures (Fig.2 (b)). In  Fig.2 (a), curves are shifted for clarity and transitions are identified.  Arrows indicate the two new transitions $\zeta_2$ and $\zeta'$ by our measurements, which encircle a new phase named phase C. Symbols representing $\zeta_1$ are taken from the previous $\rho_{zz}$ data\cite{Benoit2013}. This leads to the field-temperature phase diagram of Fig.2 (c). The phase A and B were known previously\cite{Benoit2013,Benoit2017}. In all three phases, the temperature dependence of $\rho_{xx}$ remains metallic. In phase A, $\rho_{zz}$ is insulating-like while $\rho_{xx}$ is metallic\cite{SM}; in phase B, $\rho_{zz}$ is also insulating-like while $\rho_{xx}$ is metallic, but the activation gap is smaller\cite{Benoit2013}.  In the new phase C, $\rho_{zz}$ is almost constant\cite{Benoit2013} while $\rho_{xx}$ is metallic. Interestingly, the activation gap in B-phase is smaller than that in A-phase, in contrast to the upper temperature bound of Phase C being greater than B-phase.
%

Remarkably, in the phase B, the resistivity is almost constant at T = 0.6 K.  According to our field-rotated data, in the phase B, plotted as a function of Bcos$\theta$  all $\rho_{xx}$ curves collapse on top of each other (See Fig. 3). The magnitude of the in-plane resistance (7.5 $\Omega$) combined to the thickness of the samples (24$\mu m$) yields a resistance of $500,000\Omega$ or 0.06$e^2/h$  per layer, similar to what was previously reported in the case of kish graphite\cite{Benoit2013}. The presence of an activation gap along the c-axis suggests in-plane edge transport \cite{Benoit2013,Bernevig2007} in this bulk material. In this picture, the rising of $\rho_{xx}$ with increasing field in the phase C would be due to the destruction of the c-axis activation , which ends layer-decoupling and edge transport.

\begin{figure}
\includegraphics[width=9cm]{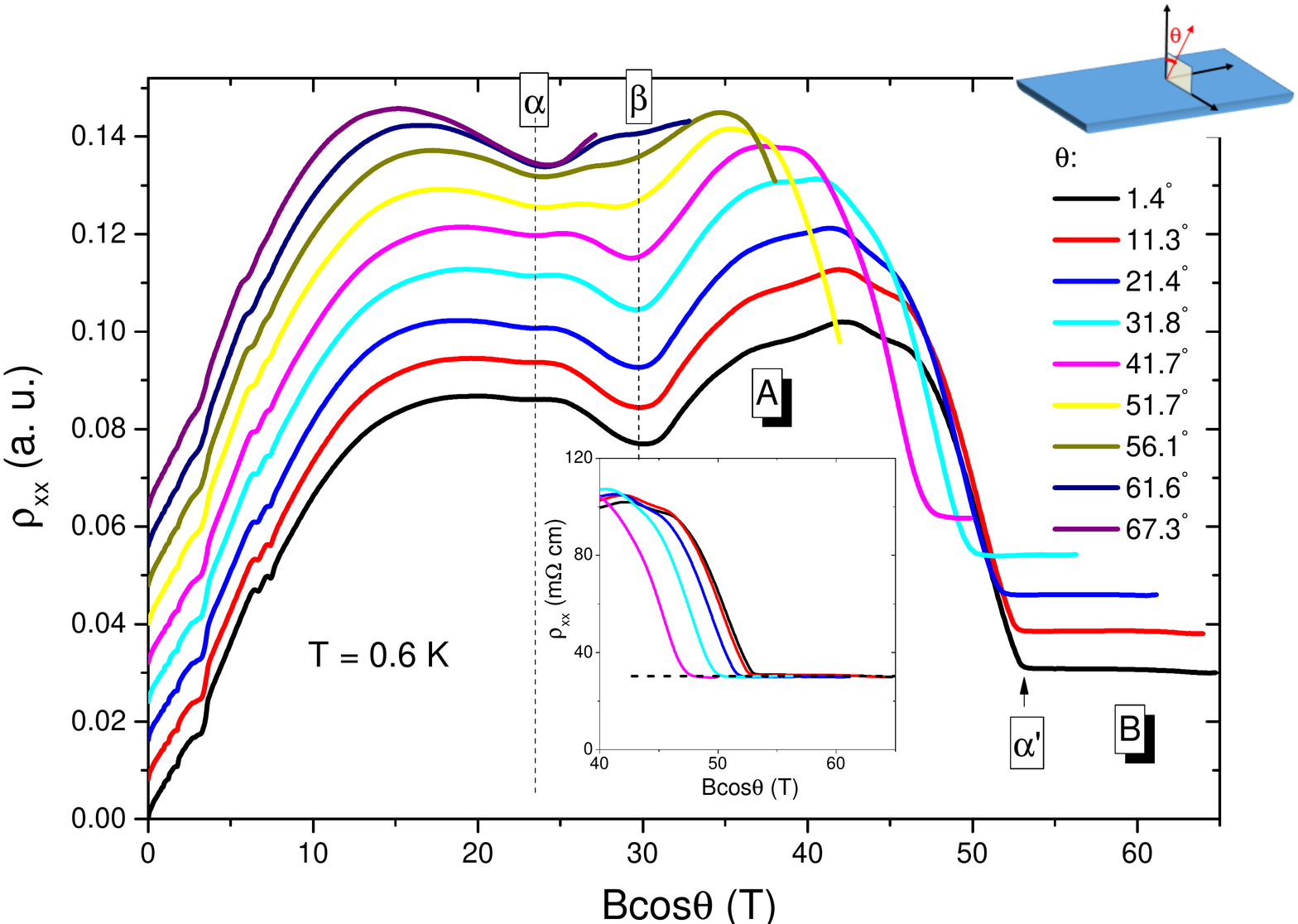}
\caption{ In-plane magnetoresistance at different angles as a function of c-axis component of the magnetic field at 0.6 K. Note that $\alpha$ and $\beta$  remain unchanged, in contrast to $\alpha'$  does.  Curves are shifted for clarity.  The inset shows the same curves with no shift in a limited field window. Note that the flat in-plane resistivity above $\alpha'$ remains unchanged. }
\end{figure}

\begin{figure}
\includegraphics[width=9cm]{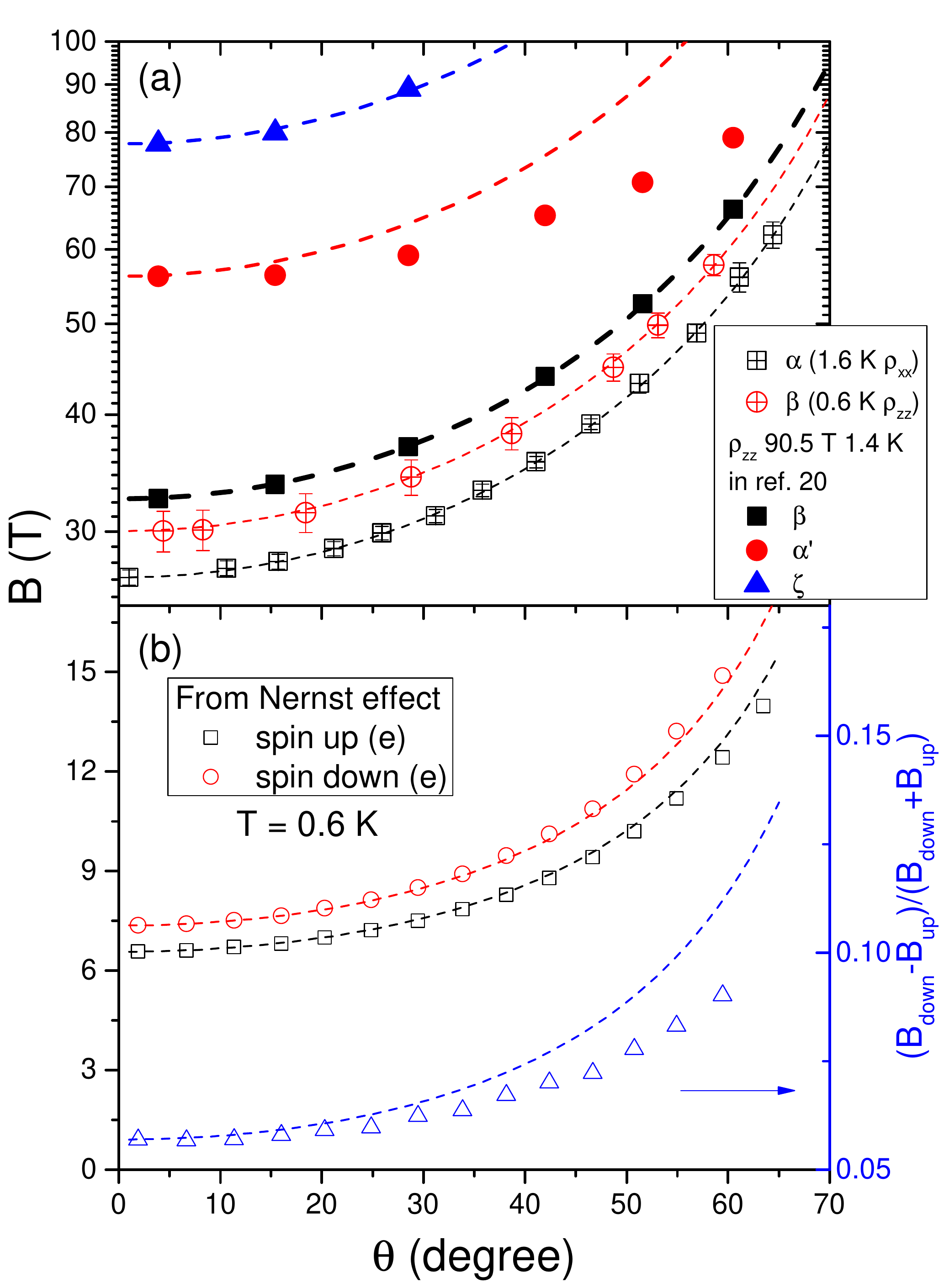}
\caption{ (a). The angle dependence of the threshold fields. Dashed lines represent a cosine dependence. Except for the $B_{\alpha'}$, all other transitions display a cosine behavior. The "+" center symbols represent anomalies in magnetoresistance data up to 65 T with regular pulse magnets\cite{SM}. The solid symbols represent anomalies in $\rho_{zz}$ up to 90.5 T\cite{ZhuEI}. (b) The angle-dependence of the Nernst peaks identified with the evacuation of spin-up and spin-down (n=1) sub-bands. Also shown is the angle dependence of the (B$_{up}$-B$_{down}$)/(B$_{up}$+B$_{down}$) ratio. Note the similarity between the behavior of this ratio which quantifies the ratio of Zeeman-to-Orbital energy and the behavior of the transition $\alpha'$ in panel a. }
\end{figure}

In graphite spin-orbit coupling is weak and band masses are light,  consequently the Zeeman energy $\frac{1}{2}g\mu_BB$ is small compared to the cyclotron energy $\hbar\omega_c$\cite{FuseyaPRL2016}. This shows up in the angle-dependent Nernst data measured at 0.6 K and up to 17 T (See Fig. S1 in the supplement). In contrast to the case of bismuth \cite{ZhuPRB2011}, the angle dependence of the Nernst peaks corresponding to the evacuation of Landau levels is smooth and close to a cosine. The Fermi surface of graphite, composed of two very anisotropic ellipsoids has a large mass anisotropy (1:7 for the electron ellipsoid, and 1:9 for the hole ellipsoid\cite{Schneider}). Because of this large anisotropy, and the smallness of the Zeeman splitting, the angular dependence of the frequency of quantum oscillations up to 60 degrees is close to a cosine fit \cite{Schneider}.

In this context, the combination of angle-dependent  $\rho_{zz}$ data at 1.4 K up to 90.5 T\cite{ZhuEI} and the angle-dependent Nernst data up to 17 T reveals a number of important features regarding the role of Zeeman and cyclotron energies in the field-induced cascade. Fig. 4 shows how the onset fields for different transitions evolve as the field is tilted off the c-axis. As seen in the Fig. 4, except the $\alpha'$ transition, all the others follow a perfect cosine behavior (See the Fig.3 and SM\cite{SM} for raw data). Interestingly, in Fig. 4(b), the ratio of Zeeman to orbital energies represented from the low field Landau level spectrum by (B$_{up}$-B$_{down}$)/(B$_{up}$+B$_{down}$) presents an angle-dependence much weaker than cosinusiodal and strikingly similar to the $\alpha'$ transition at 54 T for $\theta=0$. This indicates that the driving force behind the $\alpha'$ transition, which separates two field-induced insulating states, is the Zeeman energy.

By combining the angle-dependence and  the temperature-dependence of the threshold fields , we attempt to address three questions regarding the two phases: i)  How do electrons and holes, which pair up relate to each other? Are they the occupied and unoccupied states of same Landau sub-band  (the DW scenario) or occupied states of electron and hole sub-bands (as in the EI case)? ; ii) Is the electron-hole condensate in the weak-coupling or in the strong-coupling limit?; iii) Do pairing electrons and holes have parallel or opposite spins?  Above the QL at 7.5 T, only four sub-bands (electrons and holes, each with up and down spins) remain occupied. Phase A, relatively easy to access by regular pulsed magnetic field, was intensely studied before. Its threshold magnetic field ($\alpha$) is strongly temperature-dependent and its angular dependence follows a cosine dependence. The temperature dependence can be described by a weak-coupling BCS formula\cite{Yoshioka,Yaguchi,FauqueBook}, where the density of state (DOS) is set by the Landau level degeneracy. Its perfect cosine field-dependence (Fig. 4(b)) is to be contrasted by the angle-dependence of a Landau level, like the one evacuated at 7.5 T (Fig. 4(b)). The comparison suggests that the Zeeman energy does not play any role in driving the $\alpha$ transition. This would be true no matter the mutual orientation of spins of electrons and holes. Even with opposite spins, the Zeeman energies of pairing quasi-particle would cancel out leaving the orbital term with an exactly cosine angular dependence. The same can be said about the $\beta$ phase transition. Phase B was discovered using a 80 T magnet \cite{Benoit2013}. The $\alpha'$ transition, separating A- and B-phases has almost no temperature dependence. It is accompanied by the depopulation of at least one and most probably two Landau levels \cite{ZhuEI,Yaguchi,Akiba,Benoit2017,ArnoldCDW}. According to the present study, the angle dependence of this $\alpha'$ transition, which separates two insulators with different gaps\cite{Benoit2013} points to Zeeman energy as its driving source. One possibility is that the nature of the condensate is different in the two phases (see SM\cite{SM} for a more detailed discussion). The $\zeta$ transition field separating phase B and C has almost no temperature dependence and its angle dependence follows a cosine behaviour. Up to 90.5, both $\rho_{xx}$ and $\rho_{zz}$ remain metallic and the Hall coefficient is null\cite{ZhuEI}. As a consequence, one can  rule out the evacuation of the two remaining Landau sub-levels and the subsequent semi-metal to semi-conductor transition previously postulated\cite{ArnoldCDW}. Therefore; the field-induced destruction of the phase C cannot be attributed to any change in the Density-of-States  in the vicinity of a Landau level evacuation. If the  electrons and holes which pair up belong to different sub-bands, their different Fermi velocities would provide an opportunity for a strong magnetic field to unbind them through a purely orbital effect. An alternative scenario invoking  spin quantum fluctuations destabilizing the EI was recently proposed by Pan and et al. \cite{Pan}. Neither of these scenarios for the field-induced destruction of phase C would work in the case of a density wave.

In summary, using a very strong magnetic field, we found an additional  phase transition above 75 T and dubbed it phase C. By monitoring the angle-dependence of the multiple fields inducing phase transitions, we found that the threshold fields for all transitions, save one ($\alpha'$),  show exact cosine-dependence in angle. The in-plane magnetoresistance in phase B indicates edge transport. There is no insulating state even at 90 T and the field-induced destruction of phase B is not accompanied by the evacuation of any Landau levels. The combination of these facts leads us to conclude that the phase B is an excitonic insulator, a spin-polarized condensate of electron-hole pairs.

Z. Z. acknowledges useful discussions with Ryuichi Shindou. This work in China was supported by the 1000 Youth Talents Plan, the National Science Foundation of China (Grant No. 11574097 and No. 51861135104), The National Key Research and Development Program of China (Grant No.2016YFA0401704). Z. Z. was also supported by directors¡¯ funding grant number 20120772 at LANL. N. H. and R. D. M. acknowledges support from the USDOE BES `Science of 100T' program. The National High Magnetic Field Laboratory-PFF facility is funded by the National Science Foundation Cooperative Agreement Number DMR-1157490 and DMR-1644779
, the State of Florida and the U.S. Department of Energy. In France, this work was part of  QUANTUM LIMIT projects funded by Agence Nationale de la Recherche. B. F. acknowledges support from Jeunes Equipes de l'Institut de Physique du Coll\`{e}ge de France (JEIP). K. B. was supported by China High-end foreign expert program, 111 Program and Fonds-ESPCI-Paris.

\noindent
* \verb|zengwei.zhu@hust.edu.cn|\\

\clearpage


%
{\large\bf Supplemental Material for ``Graphite in 90 T: Evidence for Strong-coupling Excitonic Pairing''}

\setcounter{figure}{0}

\section{Angle-dependent Nernst response in low fields up to 17 T}
The Nernst effect was measured with the standard one-heater-two-thermometer method. To obtain angle-dependence of Nernst signal, we rotated the highly oriented pyrolytic graphite(HOPG) sample with a piezo-rotator. We also measured kish graphite and have similar results(not shown here). The angles were registered by two perpendicular Hall probe. The fig. S1 shows the raw data of shifted angle-dependent Nernst signal at 0.6 K. A Nernst peak correspond to the crossing of the bottom of a Landau level with the chemical potential. The (n=0,$\uparrow$) and (n=0,$\downarrow$) of the electron pocket are shown on Fig. S1. There angular dependence of peaks is reported in Fig. 4 of the main text.

\section{The anisotropy of Natural graphite}
Natural graphite samples were commercially obtained. A typical sample dimension is around $1\times$0.5$\times$0.03mm$^3$ is suitable for pulsed-magnetic-field measurements. The zero-anisotropy $\rho_{xx}/\rho_{zz}$ is around 0.002 in our samples at low temperature, shown in Fig.S2, similar to the previous report\cite{NautreAnisotropySM} and is sample dependent\cite{Benoit2013SM}.

\begin{figure}
\includegraphics[width=9cm]{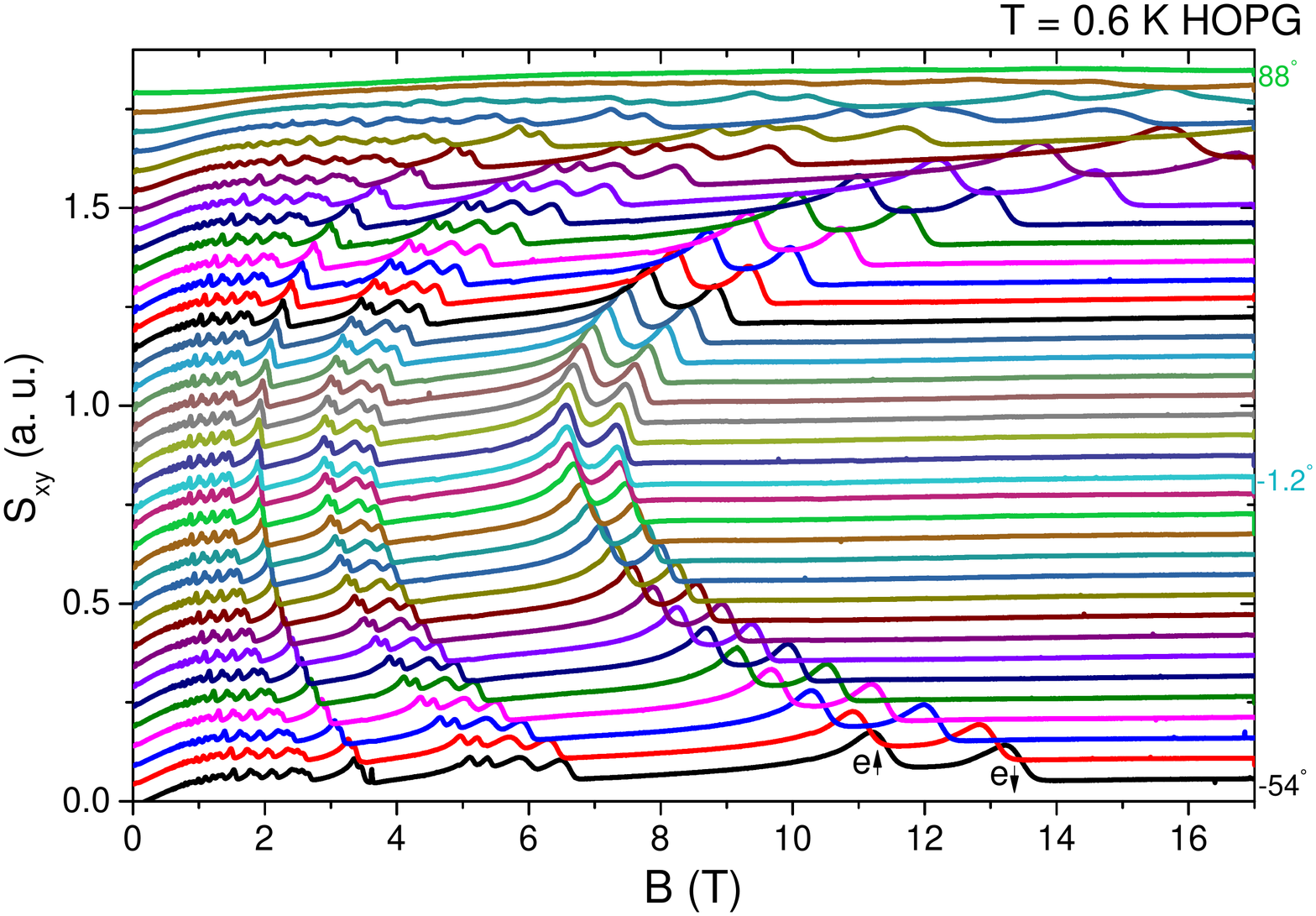}
\caption{ The field-dependence of Nernst response as the angle between field and c-axis up to 17 T at 0.6 K on a HOPG sample. The curves are shifted for clarity.}
\end{figure}
\begin{figure}
\includegraphics[width=9cm]{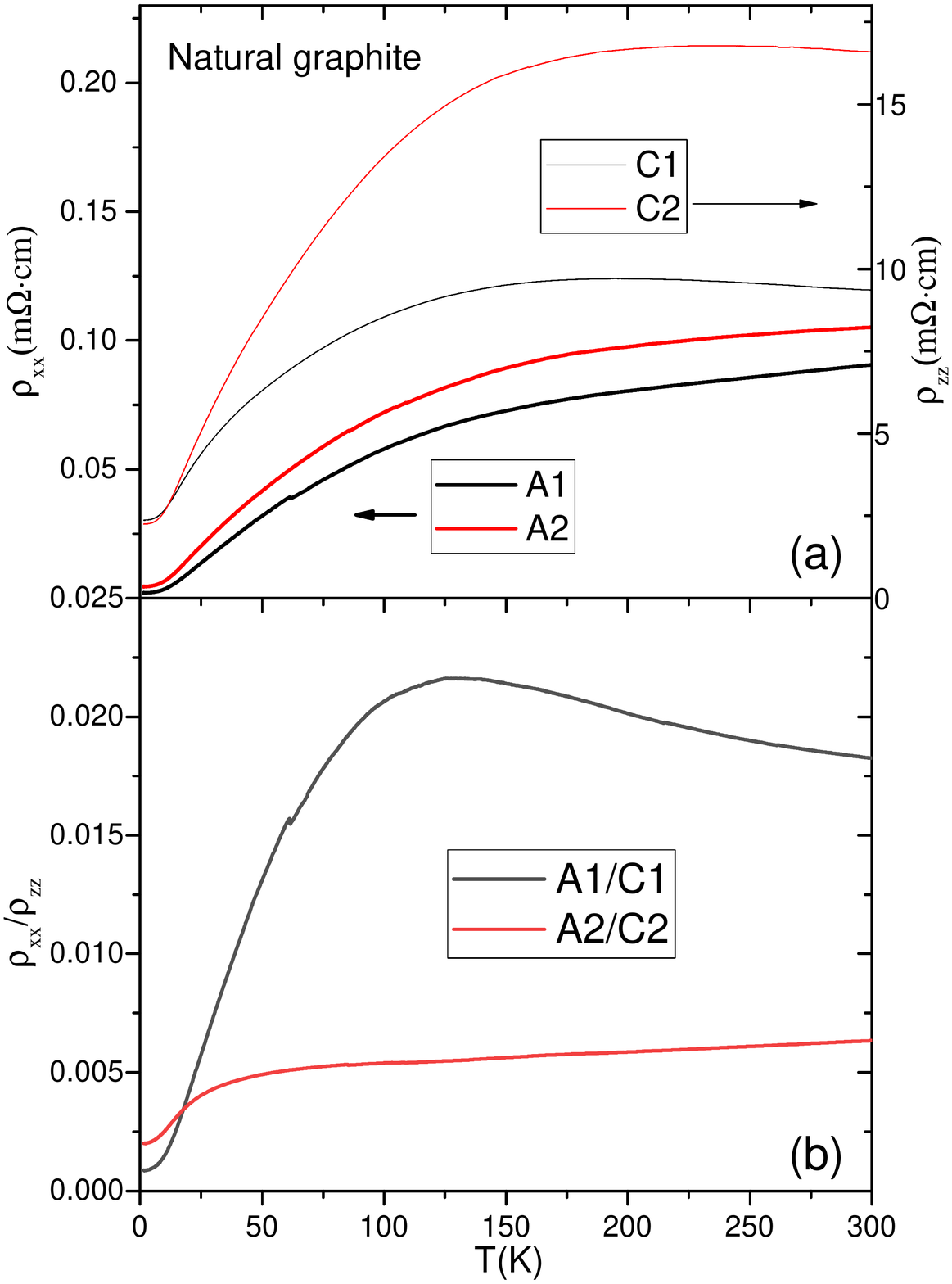}
\caption{ (a)Temperature-dependence of in-plane $\rho_{xx}$ and out-of-plane resistivity $\rho_{zz}$ in zero-field. (b) Temperature-dependence anisotropy $\rho_{xx}/\rho_{zz}$  }

\end{figure}
\begin{figure}
\includegraphics[width=9cm]{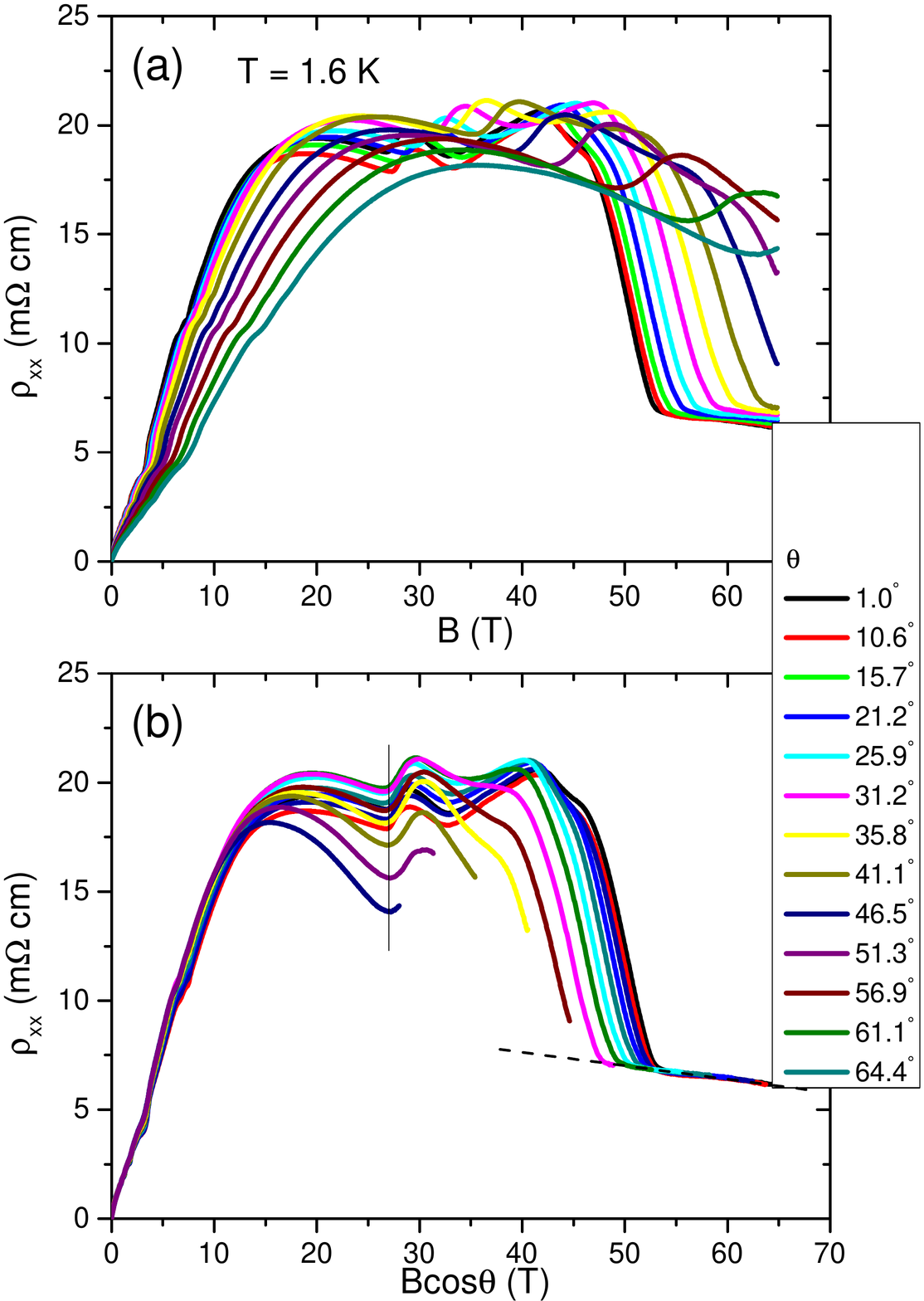}
\caption{ The field-dependence of in-plane magnetoresistance at 1.6 K at different angles up to 64 degrees (a) with normal field scale and (b) with a function of Bcos$\theta$. The dash line shows the collapse of magnetoresistance after $\alpha'$ in Bcos$\theta$. }
\end{figure}
\section{field-dependence of in-plane magnetoresistance at 1.6 K at various angles}
The Fig. S3 shows the field-dependence of in-plane magnetoresistance at 1.6 K at various angles. By plotting $\rho_{xx}$ as function of Bcos$\theta$ we find that all the curves collapse beyond $\alpha'$ transition, indicating a 2D-like behavior as observed at 0.6 K. We also checked this behavior in other temperatures and find the same trend.

\section{Second measurement on $\rho_{xx}$ and $\rho_{zz}$ up to 80 T at 1.4 K }
\begin{figure}[H]
\includegraphics[width=9cm]{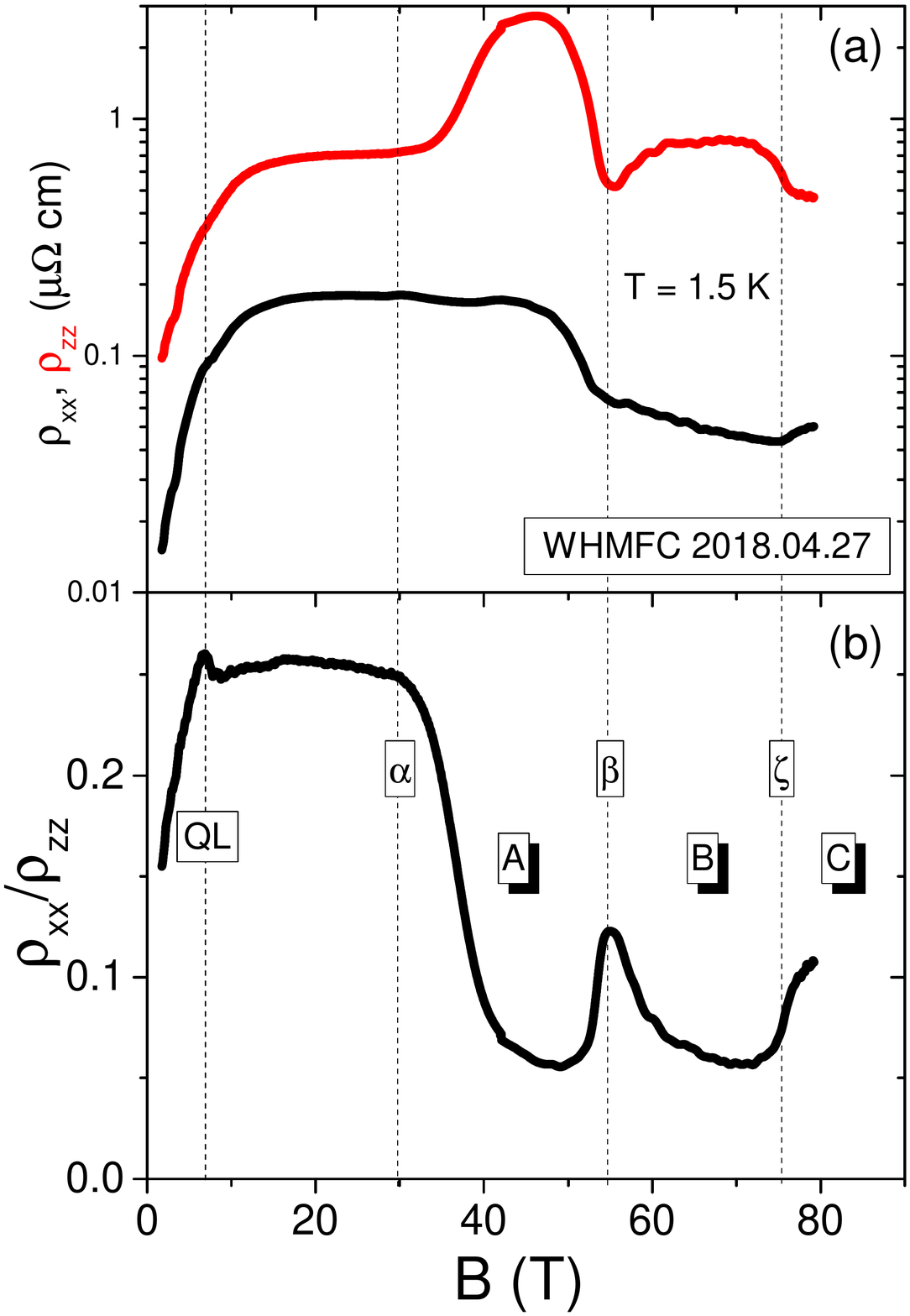}
\caption{ Field-dependence of in-plane magnetoresistance $\rho_{xx}$ and out-of-plane magnetoresistance $\rho_{zz}$ up to 80 T at 1.4 K in another two natural graphite samples. (b) The field-dependence of ratio $\rho_{xx}$/$\rho_{zz}$. The quantum limit (QL) at 7.5 T, besides the cascading field-induced phases A (28T - 54T) , B (54T - 75T)  and C (75 T - 89 T) are indexed. }
\end{figure}

Fig. S4(a) shows the field-dependence of the in-plane magnetoresistance, $\rho_{xx}$, and the out-of-plane magnetoresistance, $\rho_{zz}$, up to 80 T at 1.4 K in two others natural graphite samples. We report the ratio $\rho_{xx}/\rho_{zz}$ in Fig. S4(b) as function of the magnetic field. Like in the results reported in the main text we can identify the several phases : A (29T - 54T) , B (54T - 75T)  and C (75 T - 89 T). We note that the ratio of $\rho_{xx}/\rho_{zz}$  slightly differ from early measurement in Kish graphite \cite{Benoit2013SM} but the trend is similar. Most likely this difference is due to a difference in the density of stacking defaults between the samples.

\begin{figure}
\centering
\includegraphics[width=9cm]{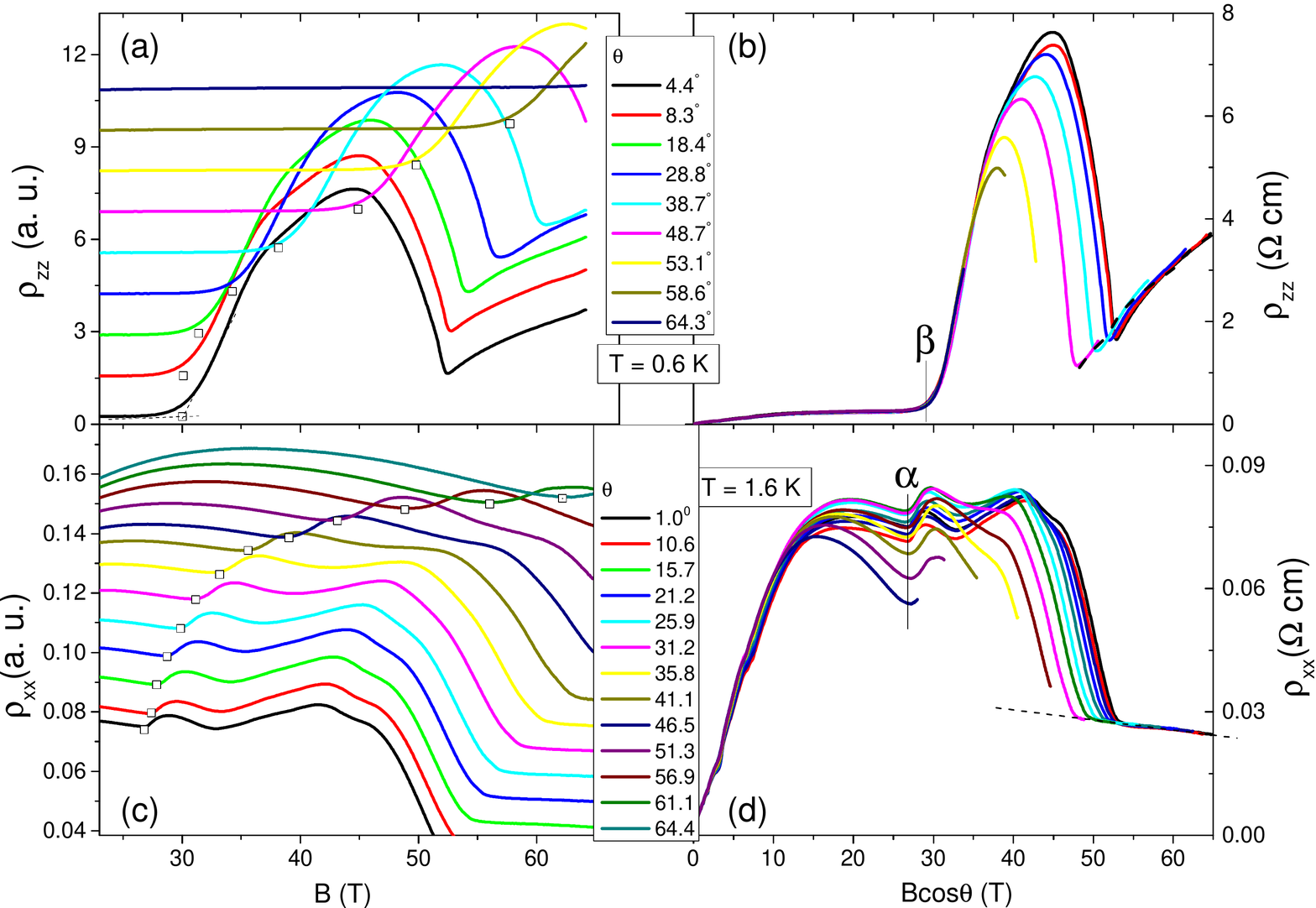}
\caption{ The field-dependence of $\rho_{zz}$ at 0.6 K in (a) normal and (b) Bcos$\theta$ scale. Interestingly, the $\rho_{zz}$ also collapse after $\alpha'$ after plotted as a function of Bcos$\theta$. The field-dependence of $\rho_{xx}$ at 1.6 K in (c) normal and (d) Bcos$\theta$ scale. The open symbols are for the transitions.}
\end{figure}
\begin{figure}
\centering
\includegraphics[width=9cm]{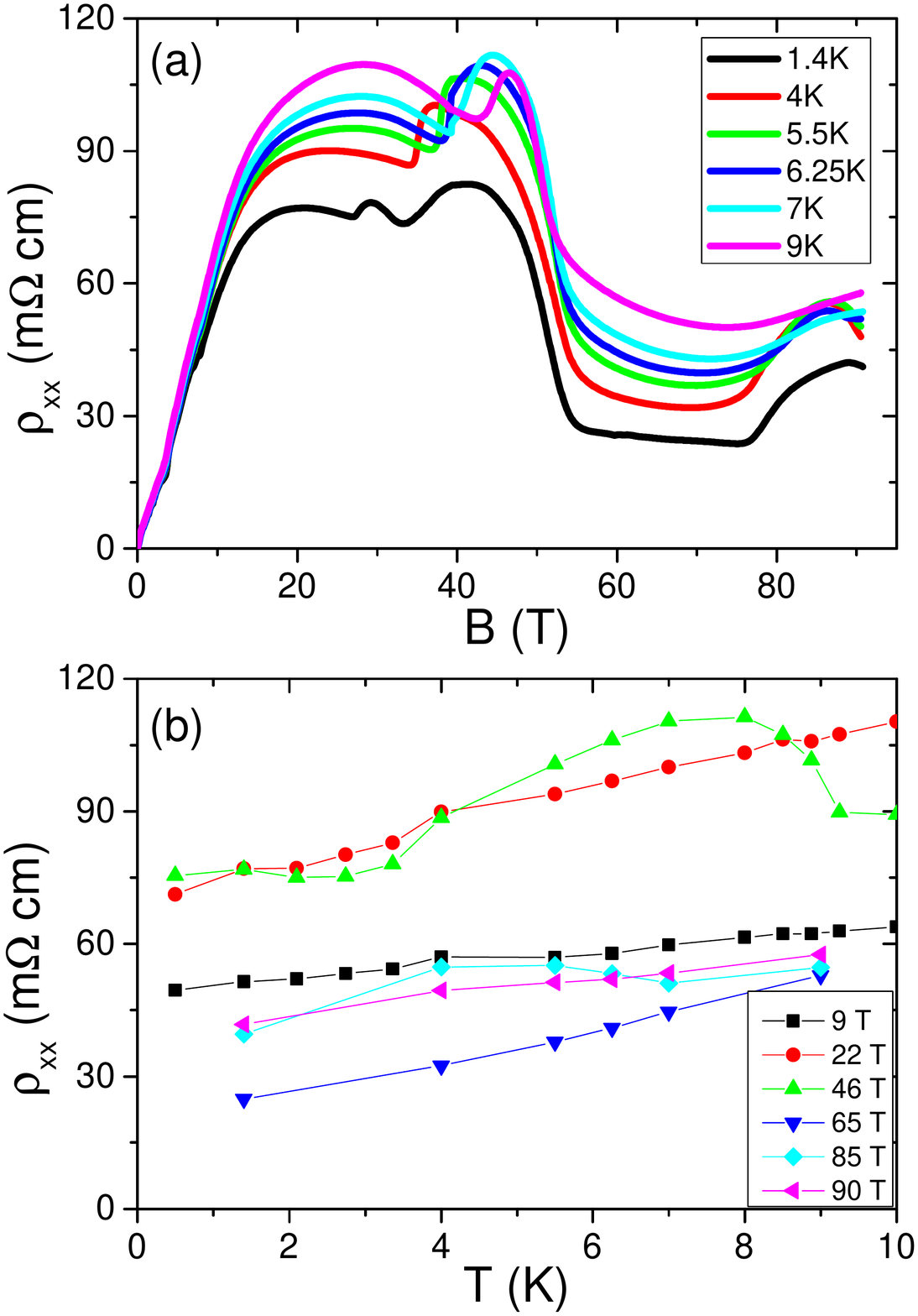}
\caption{ (a). Field-dependence of $\rho_{xx}$ at several temperatures up to 90.5 T. (b). Temperature-dependence of $\rho_{xx}$ at different fields from }
\end{figure}

The angle-dependence of $\alpha$ and $\beta$-transitions has been evaluated respectively from the measurement of $\rho_{xx}$ at T = 1.5 K (in Fig S5(a)(b)) and $\rho_{zz}$ at T = 0.6 K (in Fig S5(c)(d)) measured for a magnetic field titled from the c-axis. The threshold field $\alpha$ in $\rho_{xx}$ and $\beta$ in $\rho_{zz}$ are from the "junction" of the magnetoresistance. After plotted as a function of Bcos$\theta$, $\alpha$ and $\beta$ are same in x-axis, indicating they have exact cosine dependence in angle.

\section{Temperature-dependence of $\rho_{xx}$}
We obtain the temperature-dependence of $\rho{xx}$ from field-dependence of $\rho{xx}$ at different temperatures, shown in Fig. S6. With a regular pulsed magnet up to 65 T, we obtained more curves at various temperatures. In higher fields up to 90.5 T, we have measured at six temperatures. The data from both sets of experiments are consistent.

\section{The possible nesting vector for different phases}
\begin{figure}
\centering
\includegraphics[width=9cm]{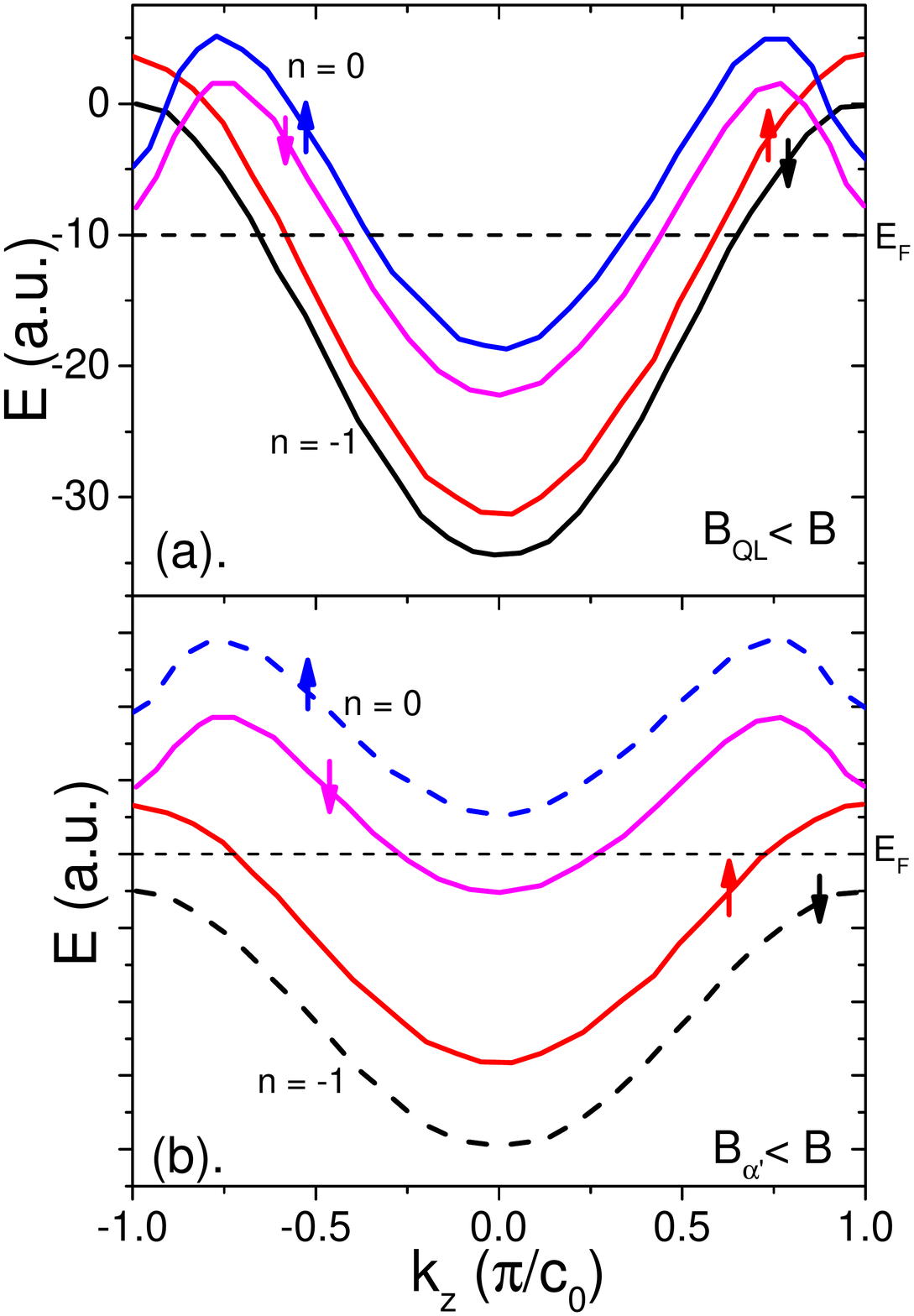}
\caption{ (a). Electronic structure of graphite under a field below $\alpha'$/peak of $\rho_{zz}$ and (b) above $\alpha'$/peak of $\rho_{zz}$ (See text for more discussions). }
\end{figure}
\begin{figure}
\centering
\includegraphics[width=9cm]{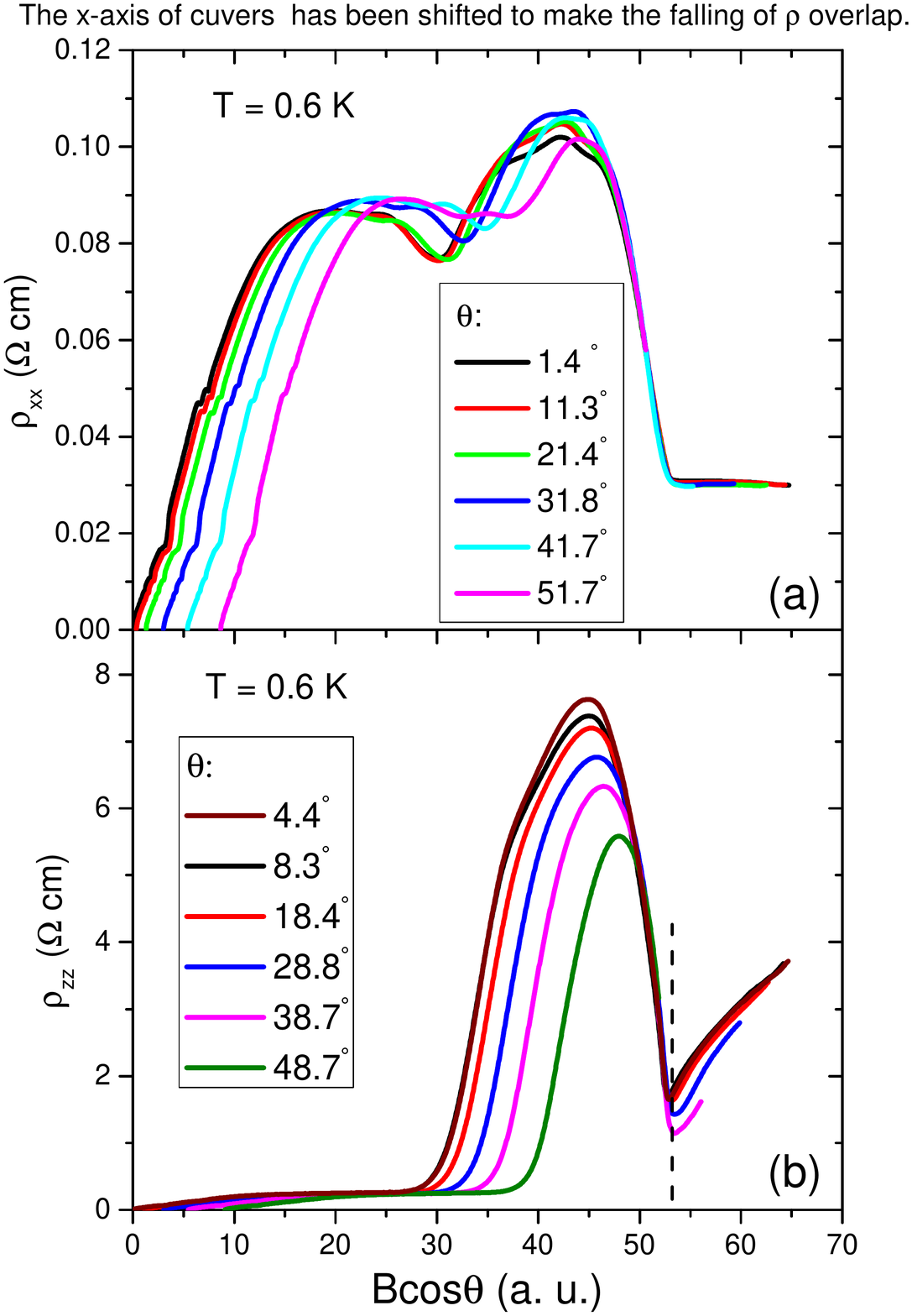}
\caption{ (a). The $\rho_{xx}$ and (b). $\rho_{zz}$ as a function of shifted fields to find the collapse the falling of the magnetoresistance curves at T = 0.6 K. This collapse indicates the same phase between the peak and the $\alpha'$. }
\end{figure}

The Fig. S7(a) and (b) show a sketch of the electronic structure of graphite under high magnetic field, before and after the two sub-bands are depopulated. The A-phase, its threshold magnetic field ($\alpha$) is strongly temperature-dependent and its angular dependence follows a cosine dependence. The temperature dependence can be described by a weak-coupling BCS formula\cite{YoshiokaSM,YaguchiSM,FauqueBookSM}, where the density of state (DOS) is set by the Landau level degeneracy. The comparison suggests that the Zeeman energy does not play any role in driving the alpha  transition. This would be true no matter the mutual orientation of spins of electrons and holes. Even with opposite spins, the Zeeman energies of pairing quasi-particle would cancel out  leaving the orbital term with an exactly cosine angular dependence.

The nature of the $\alpha'$ transition is more difficult to pin down, needs further investigations. However, we can make some speculation. After plotting in the $\rho_zz$ in a function of Bcos$\theta$, all curves collapse from the rising at $\beta$ before their downturns, seen in Fig. S5. While the curves collapse again after shifting the curves by their fields in x-axis, as seen in Fig.S8 both in $\rho_{xx}$ and $\rho_{xx}$ at 0.6 K. These behavior indicates that the peaks of $\rho_{zz}$ around 47 T are actually a sub-phase-transition which has similar angle-dependence with that of $\alpha'$, not a outcome from the competition between two phase transitions. A jump persisting in $\rho_{zz}$\cite{ZhuEISM} up to 35 K at the peak field around 47 T is another signature of the transition. As we pointed out in the previous report,\cite{ZhuEISM} this may be depopulation of two sub-bands, the band gap was assumed to open, to have a weak-coupling to strong-coupling transition. This non-cosine dependent peaks\cite{ZhuEISM} then indicates the nature of paring from two sub-bands with different angle-dependence in Zeeman energy, for instance, the pairing between (n=0,$\uparrow$) and (n=-1,$\downarrow$) become strong coupling. The $\alpha'$ is then a breakdown of strong-coupling from (n=0,$\uparrow$) and (n=-1,$\downarrow$), pairing from a band-gap, and starting of spin-polarized sub-bands (n=0,$\downarrow$) and (n=-1,$\uparrow$) pairing to have non-cosine angle dependence. Another scenario is that the peaks of $\rho_{zz}$ would be the pairing of strong-coupling EIs, to have simultaneously (n=0,$\uparrow$) with (n=-1,$\downarrow$) and (n=0,$\downarrow$) with (n=-1,$\uparrow$), which give non-cosine dependence of peaks. At the $\alpha'$, the pairing of (n=0,$\uparrow$) and (n=-1,$\downarrow$) is destroyed because two sub-bands: spin-up of electron sub-band and spin-down of hole sub-band were depopulated by magnetic fields. The destroy of nesting vector (n=0,$\uparrow$) and (n=-1,$\downarrow$) has non-cosine dependence in angle. We also note that the peak of $\rho_{zz}$ around 47 T and $\alpha'$ are almost independent on temperature, which is a feature of strong coupling EI only relying on the effective masses of the electron and hole and the dielectric constant of the material (the binding energy $E_{B}=\mu/m\epsilon^2$ (rydbergs))\cite{EIHalperinSM, EIJeromeSM} in contrast to a weak-coupling case. So the EI involves spin minority and majority bands in the upper-A phase and only spin majority bands in B-phase. The Table. I lists the summary of two scenarios. Nevertheless, We can deduce deduce an strong-coupling EI phase, pairing (n=0,$\downarrow$) with (n=-1,$\uparrow$), in the B-phase in the main text from both above-discussed scenarios.

\begin{table*}[!hbp]
\begin{tabular}{|c|c|c|c|c|}

\hline
 & T-dependence & $\theta$-dependence & Scenario 1\cite{ZhuEISM} & Scenario 2 \\
\hline
 peak of $\rho_{zz}$& weak & non-cosine  & depopulation of  (n=0,$\uparrow$) and (n=-1,$\downarrow$) & strong-coupling between e-h subband with opposite spin \\
 &  &   & weak-coupling to strong-coupling & weak-coupling to strong-coupling \\ 
\hline
 $\alpha'$& weak & non-cosine & paring across band-gap destroyed & depopulation and destroy of (n=0,$\uparrow$) and (n=-1,$\downarrow$) pairing\\
\hline

\end{tabular}
\caption{The two scenarios proposed for the peaks of $\rho_{zz}$ around 47 T and $\alpha'$ as $\theta=0$}
\end{table*}

\end{document}